\documentstyle[aps,epsf,prl,multicol]{revtex}
\begin{document}
\preprint{IASSNS-HEP-97/10}
\input epsf
%\draft
\title{The Transition Between Quantum Coherence and
Incoherence}
\author{S. P. Strong }
\address{School of Natural Sciences, Institute for Advanced Study,
Princeton, NJ 08540}
\date{25 January 1997}
\maketitle
\begin{abstract} 
We show that a transformed Caldeira-Leggett Hamiltonian
has two distinct families of fixed points,
rather than a single unique fixed point as often 
conjectured based on its connection to the anisotropic
Kondo model.
The two families are distinguished by a sharp qualitative
difference in their quantum coherence properties and
we argue that this distinction is best understood as
the result of a transition in the model between
degeneracy and non-degeneracy in the spectral function
of the ``spin-flip'' operator.
\end{abstract}
\begin{multicols}{2}

% body of paper here

The prototypical model for studying the loss of
quantum coherence is the Caldeira-Leggett or two level system model.
This model describes a two-state degree of freedom coupled
to a bath of harmonic oscillators and the Hamiltonian is
given by:
\begin{eqnarray}
\label{eq:tlsham}
H_{\rm TLS} & =  &  \Delta \sigma_x 
+ \frac{1}{2}  \sigma_z \sum_i C_i x_i \\
\nonumber & & + \sum_i  \left(\frac{1}{2} m_i \omega_i
x_i^2  + \frac{1}{2m_i} p_i^2  \right)
\end{eqnarray}
Here $C_i$ is the coupling to the $i$th oscillator,
and $m_i$, $\omega_i$, $x_i$ and $p_i$
are the mass, frequency, position
and momentum of the $i$th oscillator,
respectively.
We restrict our discussion of the model to zero
temperature and the so called
ohmic regime \cite{TLS} where the spectral density of
the bath is given by:
\begin{eqnarray}
\label{eq:ohmic}
J(\omega) &= &\frac{\pi}{2} \sum_i \frac{C_i}{m_i \omega_i}
\delta(\omega-\omega_i) \\
\nonumber
 & = & 2 \pi~ \alpha~ \omega \exp(-\omega/\omega_c)
\end{eqnarray}

In connection with questions of quantum coherence, the spin is taken
to represent a macroscopic variable and the oscillators an environmental
bath of unobservable, microscopic degrees of freedom.  If the coupling
to this bath is strong enough, then the quantum interference effects which the
isolated spin would exhibit can be wiped out by the
effects of the environment.  A quantity frequently
studied in this context
is $P(t)= \frac{1}{2} \left( 1 + \langle \sigma^z(t) \rangle \right)$. 
Here $t>0$ and
state of the system is obtained by evolving forward in time
from a $t=0$ state with $\sigma^z = 1$ and
the oscillator bath in its equilibrium state for $\sigma^z$ clamped
to $\sigma^z = 1$.  For vanishing coupling to the environment,
$P(t) = \cos 2 \Delta t$, with the oscillations resulting
from the interference between the various possible histories
of $\sigma^z(t^{\prime})$, $0 < t^{\prime} < t$.   As the coupling to 
the bath is turned on these interference effects are expected to
be gradually wiped out, representing the generic loss of
observability of interference effects between the different
possible histories of $\sigma^z(t^{\prime})$.  This corresponds
to the quantum to classical crossover for this model and
it is known that for $\alpha = \frac{1}{2}$ \cite{egger}, 
there are no oscillations of any kind and the
interference effects are completely unobservable:
$P(t) = \exp(-\Gamma t)$
with $\Gamma = \Delta^2 / \omega_c$.  

This behavior is naturally understood in a model
obtained by making a canonical
transformation \cite{TLS} on the Hamiltonian (Eq. \ref{eq:tlsham},
hereafter referred to as TLS):
$H_{\rm TLS}^{\prime} = \hat{U} H_{\rm TLS} \hat{U}^{-1}$
where:
\begin{equation}
\label{eq:xform}
\hat{U} = \exp \left(
-\frac{1}{2} \sigma_z \sum_i \frac{C_i}{m_i \omega_i^2}
\hat{p}_i \right)
\end{equation}
The new Hamiltonian takes the form:
\begin{equation}
\label{eq:tlsham2}
H_{\rm TLS}^{\prime} = \frac{1}{2} \Delta(\sigma^+ e^{-i \Omega} 
+ ~{\rm h.c.}) +
H_{\rm oscillators}
\end{equation}
where $\Omega = \sum_i \frac{C_i}{m_i \omega_i^2} p_i$.
Hereafter, we refer to Eq. \ref{eq:tlsham2} 
as the transformed Caldeira-Leggett (XCL) model.

In this model, the point $\alpha = \frac{1}{2}$ is special in that the Hamiltonian
can be converted into that of the so-called
resonant level model \cite{res_level}, which, in turn, is equivalent to
the anisotropic Kondo
problem at the Toulouse point \cite{Toulouse}.  In general the XCL model
can be connected in the limit of vanishing  $\Delta$ to the
anisotropic Kondo Hamiltonian (AKM):
\begin{eqnarray}
\label{eq:kondoham}
H_{Kondo} & = & \sum_{k,\sigma} \epsilon_k c^{\dagger}_{k,\sigma} c_{k,\sigma}
+ \frac{J_{xy}}{2} \sum_{k,q} \left( 
c^{\dagger}_{k,\uparrow} c_{q,\downarrow} S^- + ~{\rm h.c.} \right) \\
\nonumber & & +
\frac{J_{z}}{2} \sum_{k,q} \left( 
c^{\dagger}_{k,\uparrow} c_{q,\uparrow} - 
c^{\dagger}_{k,\downarrow} c_{q,\downarrow} \right) S^z
\end{eqnarray}
 via the mapping:
\begin{eqnarray}
\label{eq:mapping}
\Delta & = & \omega_c \rho J_{xy} \\
\alpha & = & \left( 1 - \frac{2}{\pi}
\arctan( \frac{\pi \rho J_z}{4} ) \right)^2
\end{eqnarray}
where $\rho$ is the density of states that follows
from $\epsilon_k$.
The AKM has
unique fixed point for all antiferromagnetic $J_z$ 
($0 < \alpha < 1$) \cite{Costi} and
it has therefore been argued that the TLS and XCL Hamiltonians
also have unique fixed points.  However, the study of
the XCL Hamiltonian by Guinea, {\it et al.}
\cite{Guinea} concluded that there was a line of fixed points 
for $0 < \alpha < \frac{1}{2}$ and that only for 
$\alpha ~^>_{\_} \frac{1}{2}$, was the system described by the
unique Kondo fixed point.  It would have great intuitive
appeal if the ``incoherent'' and ``coherent'' phases
of the TLS model corresponded to different
fixed points, however, a recent study \cite{Sudip} of the long time
behavior 
of $\langle \sigma^z(t) \sigma^z(0) \rangle $ 
concluded that that asymptotic behavior is $\sim t^{-2}$
for any $\alpha$ with $0 <  \alpha < 1$. This result is
consistent with a unique fixed point.
Thus, while there is some
uncertainty, the accepted wisdom is that 
all three Hamiltonians exhibit only a single fixed point
and that, consequently, no true, long time 
distinction exists between the ``coherent'' and ``incoherent''
phases.
The purpose of this
paper is to demonstrate that, while the anisotropic Kondo
model may possess a unique fixed point for antiferromagnetic
$J_z$, and we find no evidence for multiple fixed points
in the regime $0 < \alpha < 1$ for the TLS Hamiltonian, the 
XCL Hamiltonian {\it does possess} an entire family of fixed points
distinguished by different values of $\alpha$.  Further,
there is a true, qualitative, long time distinction between the behavior
for $0 < \alpha < \frac{1}{2}$ and
$ \frac{1}{2} < \alpha < 1$.  This
distinction is directly related to questions of quantum (in)coherence 
and, we believe, to the effective transition between
degeneracy and non-degeneracy 
of the action of the spin flip operator.

As mentioned,  questions of quantum coherence have traditionally focused
on the quantity $P(t)$ defined above and the question of
whether or not it exhibits oscillations at ``intermediate 
times'' (times of order the Kondo scale).  In the XCL model, there
exists a more direct probe of quantum coherence: the off-diagonal components
of the density matrix describing the two-state degree of freedom
when the oscillator bath is traced over \cite{zurek,PTdiag}.
The sum of the two 
off-diagonal components of the density matrix is
given by $\langle \sigma^x \rangle$, so we 
choose as our probes of quantum coherence
the correlation functions of $\sigma^x$.

First, consider the behavior of the time symmetrized correlation:
$F(t) = \frac{1}{2} \langle \{ \sigma^x(t) ,\sigma^x(0) \} \rangle$
in the solvable limits: $\alpha = 0$ and $\alpha = 1/2$.  At
$\alpha = 0$, the problem is trivial and 
$F(t) = 1 $
because $\langle \sigma^x \rangle = 1 $; the system
has maximal quantum coherence in the
sense that the off-diagonal components of the
density matrix are as large as the diagonal.  At
$\alpha = 1/2$, the Toulouse re-fermionization
may be used and 
$F(t) = \exp(-\Gamma t)$,
identical to $2\left( P(t)-\frac{1}{2} \right)$.  At this
point there is no sign of any quantum coherence:
not only is $\langle \sigma^x \rangle = 0$,
but the correlations of $\sigma^x$ entering
into $F(t)$ decay faster than
any powr of $t$.

Now consider other values of $\alpha$;
here we follow the numerical approach of
\cite{Sudip} and study the imaginary time correlation function
$\langle \sigma^x(\tau) \sigma^x(0) \rangle$ using
the Coulomb gas (CG) language \cite{phil}. 
Recall
that the CG model related to the
AKM is a one dimensional model with alternating plus and minus
charges which interact with a logarithmic
Coulomb interaction whose strength is 
proportional to $\alpha$\cite{phil}.
We choose 
to use an inverse squared Ising model (ISI) as a
specific realization of the CG with a short distance regularization
provided by the lattice.
The ISI is defined on $N$ sites with the Hamiltonian:
\begin{eqnarray}
\label{eq:ISI}
H_I & = & - \frac{J_{NN}}{2} \sum_{0^<_{\_}i<N} S_i S_{i+1} - \frac{J_{LR}}{2}
\sum_{i < j} \frac{(\pi/N)^2 S_i S_j}{\sin^2[\pi(j-i)/N]}
\end{eqnarray}
where $J_{LR} = \alpha$,
$N  =  \beta_{XCL}$ and
$\Delta  \sim  2 \exp \left( - J_{NN} - J_{LR}(1 + \gamma) \right)$,
with $\gamma$ Euler's constant.

The correlation function 
${\cal G}(\tau) = \langle  \sigma^+(\tau) \sigma^-(0)  \rangle $
is given by the ratio of a certain ``twisted'' CG partition
function to the usual partition function.  
The twisted partition function is defined by restricting 
the sum over states to those where the charges on either side $x=0$
are positive and those on either side of  $x=\tau$ are negative,
and requiring that,
elsewhere, the charges alternate as usual. 
This ratio can be computed in the ISI language 
as the inverse of the usual ISI partition function
times the sum over all states which have $\sigma^z(0) = 1$
and $\sigma^z (\tau) = 1$ of a Boltzman like term, $\exp( - E_{modified})$.
$E_{modified}$ is defined by:
$ E_{modified}  =  2E^{\prime} + 2E^{\prime\prime} - E - 3E^0 $.
$E$, $E^{\prime}$
$E^{\prime\prime}$
and $E^0$
are the energies computed using the ISI Hamiltonian
and, respectively, the actual configuration considered, the configuration
obtained from the actual by removing all domain walls from spins
$1$ to $\tau$, the configuration
obtained from the actual by removing all domain walls from spins
$\tau+1$ to $N$ and the fully polarized configuration.
In the CG language, this reverses the signs of all the
charges between $x=0$ and $x=\tau$, the appropriate
operation for the change in allowed configurations
induced by $\sigma^+(\tau)\sigma^-(0)$.

\begin{figure}
\narrowtext
\centerline{ \epsfxsize = 3in
\epsffile{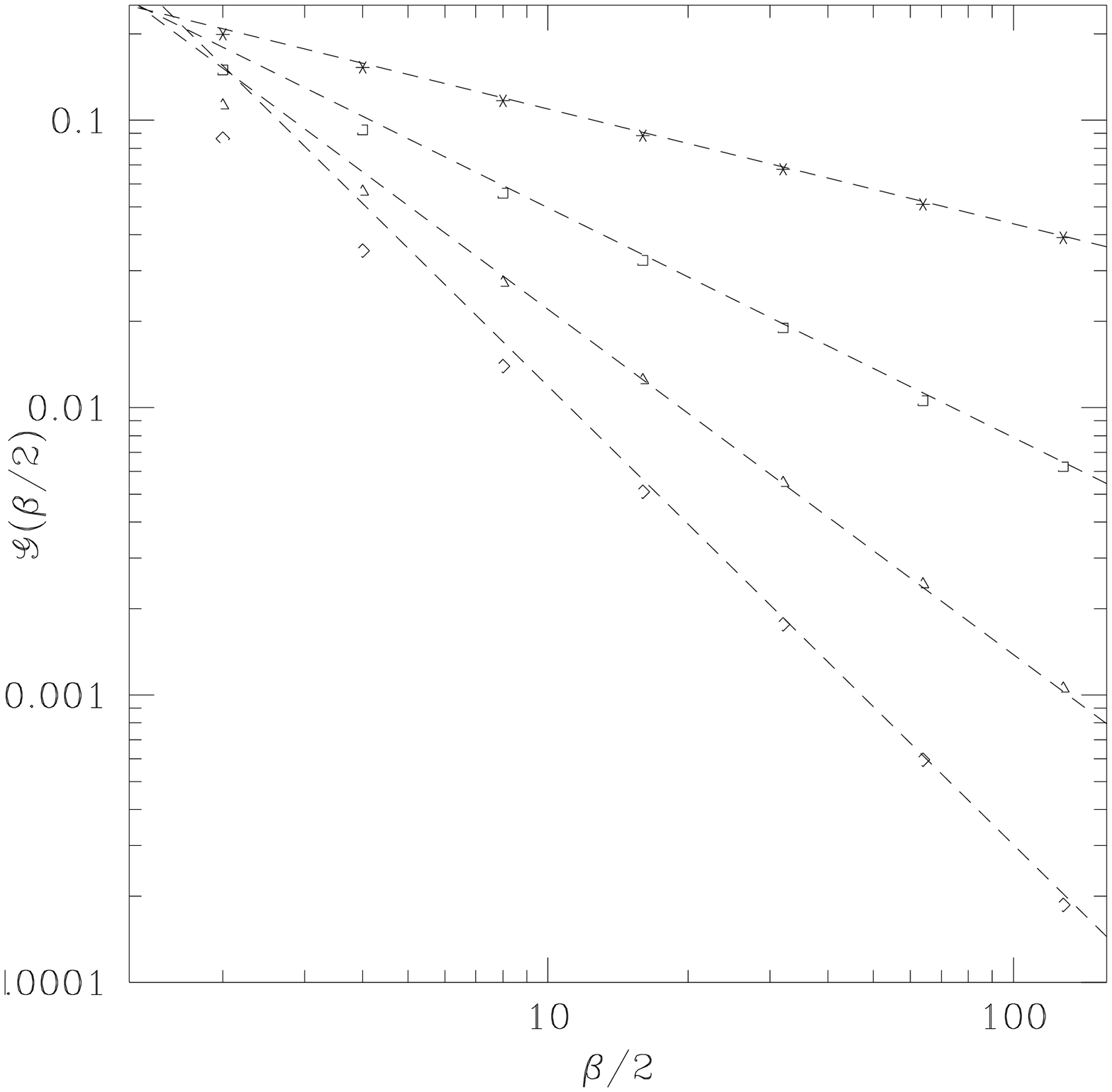}}
\caption{Log-log plot of ${\cal G}(\tau = \beta/2 = N/2)$
versus $\beta / 2 = N/2$.
$\ast$ correspond to
$\alpha = .2$, $T_K$ (as defined in Ref. 7)
$\sim 0.70$. $\Box$ to $\alpha = .4$,
$T_K \sim .66$. $\triangle$ to $\alpha = .6$,
$T_K \sim .62$. $\Diamond$ to $\alpha = .8$,
$T_K \sim .62$.  Dashed lines are guides to the eye
with slopes of .4, .8, 1.2 and 1.6: the expected behaviors
if ${\cal G}(\tau = \beta /2 = N/2) \propto \tau^{-2 \alpha}$.}
\label{fig:halfsep}
\end{figure}

\begin{figure}
\narrowtext
\centerline{ \epsfxsize = 3in
\epsffile{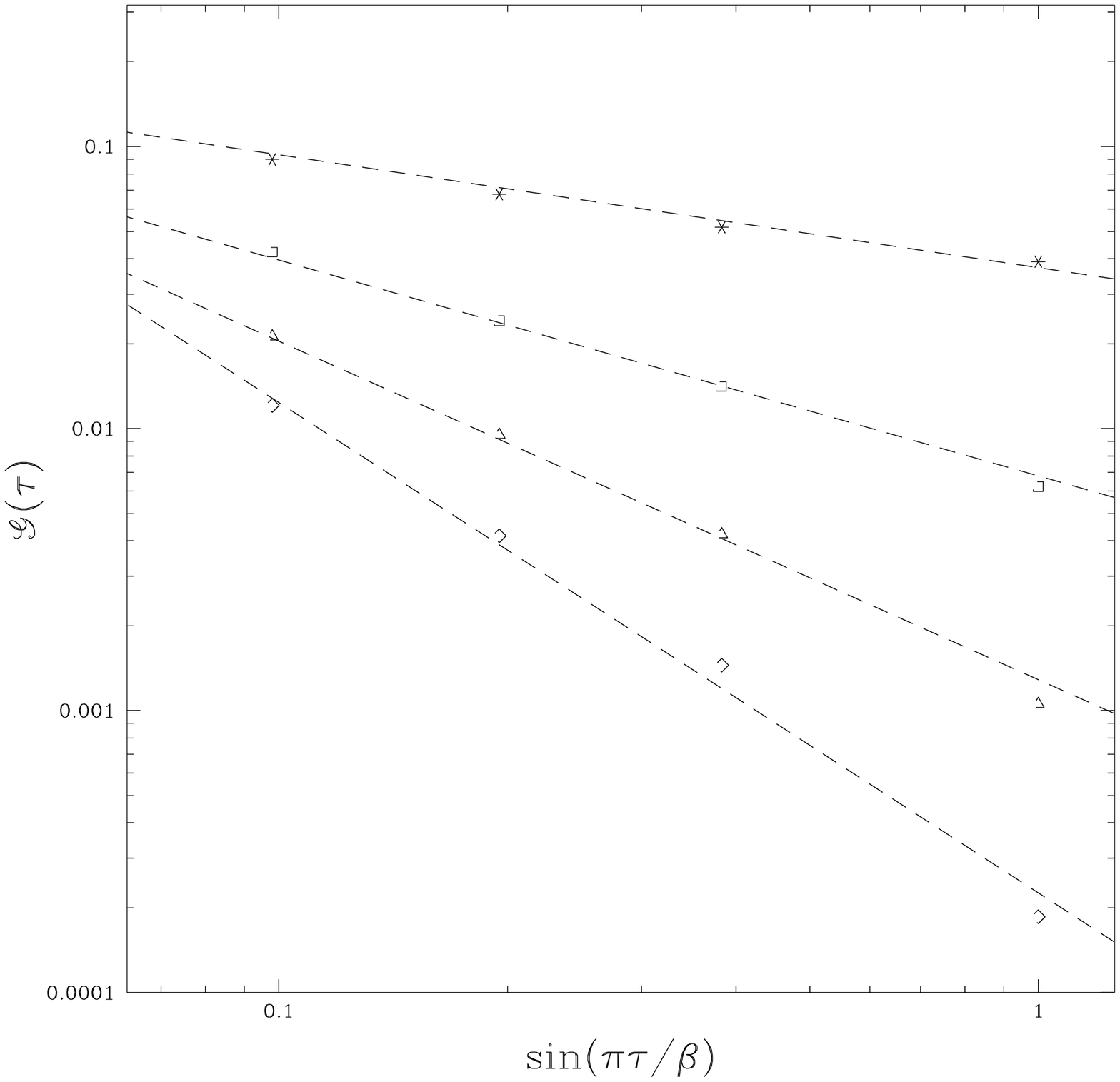}}
\caption{Log-log plot of ${\cal G}(\tau)$
versus $\sin\left(\pi\tau / \beta\right)$
for $N = \beta = 256$.
$\ast$ correspond to
$\alpha = .2$, $T_K \sim 0.70$. $\Box$ to $\alpha = .4$,
$T_K \sim .66$. $\triangle$ to $\alpha = .6$,
$T_K \sim .62$. $\Diamond$ to $\alpha = .8$,
$T_K \sim .62$.  Dashed lines are guides to the eye
with slopes of .4, .8, 1.2 and 1.6: the expected behaviors
if ${\cal G}(\tau) \propto \sin^{-2 \alpha} \left( \pi\tau / \beta \right)$.}
\label{fig:diffsep}
\end{figure}

We have used Monte Carlo methods
to study 
${\cal G}(\tau)$
and typical results 
are depicted in Figs. \ref{fig:halfsep} and \ref{fig:diffsep}.
\vspace*{-0.6cm}
\vspace*{-0.6cm}
For distances large compared to the Kondo scale, the
results are clearly well described by a power decay of
${\cal G}(\tau)$ as $\tau^{-2 \alpha}$. The long time
behavior of $\cal G$ is therefore different for
different $\alpha$, which establishes that the XCL model possesses a line
of fixed points, not a single strong coupling fixed point \cite{caveat}.
The result can be simply understood in the CG language where the insertion
of a $\sigma^+$($\sigma^-$) acts to change the allowed configurations
to those requiring two consecutive positive(negative) charges about
the insertion point.  The finite fugacity of other charges 
translates this into an effective charge insertion for
distances long compared to the Kondo scale, and
this insertion of an
extra charge can not be screened
since plus and minus charges are required to alternate
away from insertions.  Therefore, at long distances, $G$ behaves as 
the charge insertion correlator with unrenormalized 
$\alpha$.

While all $\alpha$ correspond to different fixed points, there is
an important distinction between $\alpha < \frac{1}{2}$ and
$\alpha > \frac{1}{2}$.
For $0 < \alpha < \frac{1}{2}$, the long imaginary time
behavior of ${\cal G}$ dominates the low frequency behavior
and a continuation from Matsubara frequencies to real
frequencies will give a low frequency singularity in
$G_{ret}(\omega) \sim |\omega|^{-1+2\alpha} 
e^{ i \frac{\pi}{2}  (1-2\alpha) {\rm sgn}(\omega)}$.
Since the spectral
function diverges at low frequencies, we know that 
$G(t) \sim  F(t) \sim  t^{-2 \alpha}$, 
with the prefactor in $F$ vanishing
as $\alpha \rightarrow \frac{1}{2}$ \cite{Gnote}.  Throughout this region,
the susceptibility of the systems with respect to
a perturbation coupling to $\sigma^x$:
\begin{equation}
\chi_{\rm coh} = \int_0^{\infty} dt ~[\sigma^x(t),\sigma^x(0)]
\end{equation}
is {\it divergent}; the system is enormously sensitive 
to any small perturbation
that tends to induce coherence as defined by finite off-diagonal elements
in the density matrix. Conversely, for
$ \frac{1}{2} < \alpha < 1$, $\chi_{\rm coh}$ is finite
because of the rapid decay of the correlation functions of
$\sigma^x$, and this sensitivity is absent. 

We believe that the disappearance of the divergent
susceptibility may be interpreted as a ``transition'' from
degeneracy to non-degeneracy in the action of $\Delta$,
as has been previously suggested
\cite{usspectral}.  To understand this, consider the meaning
of $\sigma^+$($\sigma^-$) in the XCL language.  The Hilbert space of the
system is spanned by the $\Delta = 0$ eigenstates which consist
of two towers of oscillator eigenstates
($|A_i \rangle$ and $|B_i \rangle$). The towers are
distinguished by the value of $\sigma^z$, but are otherwise identical
($|A_i \rangle  = \sigma^+ |B_i \rangle$).
At finite $\Delta$ and for $\alpha < 1$, the groundstate
is a complicated superposition of these states
($|\Psi_0 \rangle = \sum_i 
\left( \lambda_i^a |A_i \rangle + \lambda_i^b |B_i \rangle \right)$)
with equal weight
from coming from each tower 
($ \sum_i |\lambda_i^a|^2 = \sum_i |\lambda_i^b|^2 = \frac{1}{2}$).  
$\sigma^x$ converts the 
$\lambda_i^b$'s into
the $\lambda_i^a$'s (and vice versa),
probing the phase relationship between the 
$\lambda^a$'s and the $\lambda^b$'s.  The vanishing of
$\langle \sigma^+ \rangle$ implies that the phases of
the  $\lambda^a$'s and the $\lambda^b$'s are, on average,
completely uncorrelated in the ground state.  
This might appear
natural since, away from
$\alpha = 0$, the matrix elements of the $\Delta$ term
in the Hamiltonian connecting $|A_i\rangle$ and $|B_i\rangle$
vanish in the $\Delta = 0$ groundstate due to an orthogonality
catastrophe.  However, consider
the spectral function in the 
$\Delta = 0$ groundstate for $e^{i\Omega}$:
\begin{eqnarray}
\label{eq:TLSrho}
\rho_{\Delta}(\omega) & = & \sum_m |\langle m | e^{i \Omega} | GS \rangle |^2
\delta(\omega - E_m) \\
\nonumber & = &
\Gamma^{-1}(2 \alpha)~\theta_+(\omega)~
\omega^{-1+2 \alpha} \omega_c^{-2 \alpha}
\exp(-\omega / \omega_c)
\end{eqnarray}
For small $\alpha$ the spectral function is strongly peaked about
small energy.  The phases of the $\lambda_i^a$ 
associated to various low lying states, $|A_i\rangle$,
should therefore
be  weakly correlated
with the phases of a large number of $\lambda_j^b$ which 
represent states nearly
degenerate in energy. The dephasing of these states resulting
from their energy difference with respect to the
$\Delta = 0$ Hamiltonian is very slow and,
since the phases of each of these $\lambda_j^b$
should be correlated with the phases of a large number of $\lambda_k^a$,
which again represent states nearly
degenerate in energy (and nearly degenerate with $|A_i\rangle$ ),
there should be an increasingly strong
tendency for the phases of the $\lambda^a$'s and the $\lambda^b$'s
to correlate in the limit of small $\alpha$.  It is 
this near degeneracy of the 
perturbation theory in $\Delta$ which 
underlies the very slow decay of the $\sigma^x$ correlations
in time and the diverging susceptibility to coherence,
$\chi_{\rm coh}$.
As we tune $\alpha$ up from $0$, we move away from
the case where $\Delta$ couples completely or even
nearly
degenerate states, until at $\alpha = \frac{1}{2}$, the
spectral function, $\rho_{\Delta}$, is completely flat, 
perturbation
theory in $\Delta$ is non-degenerate and $\chi_{\rm coh}$ is finite.

Since evolution of the effects of $\Delta$ in the XCL from degenerate
to non-degenerate is
not merely quantitative but has qualitative changes in the long time
behavior of the system associated with it, it is 
natural to conjecture
that these changes underlies the evolution from coherence
to incoherence of the TLS model.
In fact, 
the near degeneracy with respect to the $\Delta=0$
Hamiltonian of the states connected by $\Delta$
is also exactly
what is required allow quantum interference to
be observable as oscillations in $P(t)$ \cite{usspectral}
and
the oscillations in $P(t)$
vanish at $\alpha = \frac{1}{2}$ \cite{egger},
precisely 
the point where the susceptibility to ``coherence'' became finite.

Given this, it is natural to ask whether or not the TLS model
has a unique fixed point for $0< \alpha <1$ or several different fixed points,
some of which exhibit quantum coherence and some of which don't.
If one takes the $\sigma^x$ operator whose correlations
distinguished the fixed points of the XCL model and maps it
back to the TLS model, it becomes the operator 
$\frac{1}{2}\left(\sigma^+ e^{i \Omega} + ~{\rm h.c.} \right)$,
so that the fact that it exhibits an $\alpha$ dependent power
law in its correlation functions is unsurprising and not 
necessarily indicative
of multiple fixed points.  Likewise, the off-diagonal elements
of the density matrix for the spin of the TLS model have
no interesting $\alpha$ dependence, since they are finite
for finite $\Delta$ and any $\alpha < 1$.  It therefore appears
very likely that the TLS problem for $0 < \alpha < 1$ is
described by a single unique fixed point, although the interesting
change in the intermediate time properties of $P(t)$
appears intimately connected to the 
different quantum coherence properties of the different XCL fixed points.

Similarly to the TLS case, for the AKM, the operators which 
correspond to the $\sigma^x$ operator of the XCL take the form
$\frac{1}{2}\left(\sigma^+ 
e^{-i (\sqrt{1-\alpha}) \phi(0)} + ~{\rm h.c.} \right)$
and again the $\alpha$ dependent correlation functions neither
require nor suggest multiple fixed points.  Thus, the results
presented here are not in contradiction to previously known results.
They do demonstrate the surprising fact that the renormalization 
group flows for the XCL model are vertical in $(\alpha,\Delta)$ space
for $0 < \alpha < 1$,
with $\Delta$ growing large, but $\alpha$, as measured by
the correlations of $\sigma^x$ unrenormalized at large scales. 
This result is of particular interest since a number of
problems 
involving coupled Luttinger liquids have been connected to the
AKM because they are
analogous to the XCL model.  These
models may well {\it not} exhibit a unique fixed point
as has commonly been supposed based on that connection. 

In conclusion, we have studied the correlation functions
of the $\sigma^x$ operator of the transformed Caldeira-Leggett
model defined
by Eq. \ref{eq:tlsham2}, and we find clear evidence for
two distinct families of fixed points.  The two families
are distinguished by the (in)finiteness of a particular
susceptibility which is closely connected to the question
of the quantum coherence of the $\sigma$ variable.  We identify
the transition in behavior between the two regimes as an
effective ``transition'' from degenerate to non-degenerate
action by the operator $\Delta$ and have shown that it is closely connected
to the quantum to classical crossover of the Caldeira-Leggett
model.

S. P. S. acknowledges useful conversations with David G. Clarke
and financial support from DOE grant DE-FG02-90ER40542.

% now the references. delete or change fake bibitem. delete next three
%   lines and directly read in your .bbl file if you use bibtex.

\end{multicols}

% \begin{table}
% \caption{}
% \label{}
% \end{tabular}
% \end{table}


\begin{references}
\bibitem{PTdiag}    $P(t)$
measures the time dependence of the diagonal elements of
this density matrix under a particular set of circumstances
and is a less direct probe of quantum coherence.  However, 
it is much more closely related to
possible experimental realizations and enjoys other advantages
for the TLS model.
\bibitem{caveat} At least the physics is not governed by a
unique ``Kondo'' fixed point since the power law behavior sets
in on scales long compared to the Kondo scale. There is no reason to
believe that at some much longer scale there is crossover to an
$\alpha$ independent behavior, but this can not be ruled out
numerically.
\end{references}
\end{document}